\documentclass[twocolumn,natbib,iop,hyperref]{aastex62}
\usepackage{natbib}

\bibpunct{(}{)}{;}{a}{}{,}
\usepackage{multirow}
\usepackage{color}
\usepackage[varg]{txfonts}

\newcommand{\cgs}{erg s$^{-1}$ cm$^{-2}$}

\newcommand{\lya}{Lyman-$\alpha$ }
\newcommand{\lyans}{Lyman-$\alpha$}
\submitjournal{The Astrophysical Journal Letters}

\accepted{August 29, 2018}
\shorttitle{MUSE Spectroscopic Identifications of Ultra-Faint Emission Line Galaxies}
\shortauthors{Maseda et al.}

\begin{document}

\title{MUSE Spectroscopic Identifications of Ultra-Faint Emission Line Galaxies with M$_{\mathrm{UV}}\sim$ -15\footnote{Based on observations made with ESO telescopes at the La Silla Paranal Observatory under program IDs 094.A-2089(B), 095.A-0010(A), 096.A-0045(A), and 096.A-0045(B); and based on data obtained with the NASA/ESA \textit{Hubble Space Telescope} which is operated by the Association of Universities for Research in Astronomy, Inc., under NASA contract NAS 5-26555.}}
\correspondingauthor{Michael V. Maseda}
\email{maseda@strw.leidenuniv.nl}

\author[0000-0003-0695-4414]{Michael V. Maseda}
\altaffiliation{NOVA Fellow}
\affil{Leiden Observatory, Leiden University, P.O. Box 9513, 2300 RA, Leiden, The Netherlands}
\author{Roland Bacon}
\affil{Univ Lyon, Univ Lyon1, Ens de Lyon, CNRS, Centre de Recherche Astrophysique de Lyon UMR5574, 69230, Saint-Genis-Laval, France}
\author[0000-0002-8871-3026]{Marijn Franx}
\affil{Leiden Observatory, Leiden University, P.O. Box 9513, 2300 RA, Leiden, The Netherlands}
\author[0000-0003-4359-8797]{Jarle Brinchmann}
\affil{Leiden Observatory, Leiden University, P.O. Box 9513, 2300 RA, Leiden, The Netherlands}
\affil{Instituto de Astrof{\'\i}sica e Ci{\^e}ncias do Espa\c{c}o, Universidade do Porto, CAUP, Rua das Estrelas, PT4150-762 Porto, Portugal}
\author[0000-0002-0668-5560]{Joop Schaye}
\affil{Leiden Observatory, Leiden University, P.O. Box 9513, 2300 RA, Leiden, The Netherlands}
\author[0000-0002-3952-8588]{Leindert A. Boogaard}
\affil{Leiden Observatory, Leiden University, P.O. Box 9513, 2300 RA, Leiden, The Netherlands}
\author[0000-0003-0068-9920]{Nicolas Bouch\'e}
\affil{Institut de Recherche en Astrophysique et Plan\'etologie (IRAP), Universit\'e de Toulouse, CNRS, UPS, F-31400 Toulouse, France}
\affil{Univ Lyon, Univ Lyon1, Ens de Lyon, CNRS, Centre de Recherche Astrophysique de Lyon UMR5574, 69230, Saint-Genis-Laval, France}
\author[0000-0002-4989-2471]{Rychard J. Bouwens}
\affil{Leiden Observatory, Leiden University, P.O. Box 9513, 2300 RA, Leiden, The Netherlands}
\author{Sebastiano Cantalupo}
\affil{ETH Z\"urich, Department of Physics, Wolfgang-Pauli-Str. 27, 8093 Z\"urich, Switzerland}
\author[0000-0003-0275-938X]{Thierry Contini}
\affil{Institut de Recherche en Astrophysique et Plan\'etologie (IRAP), Universit\'e de Toulouse, CNRS, UPS, F-31400 Toulouse, France}
\author[0000-0002-0898-4038]{Takuya Hashimoto}
\affil{Department of Environmental Science and Technology, Faculty of Design Technology, Osaka Sangyo University, 3-1-1, Nagaito, Daito, Osaka 574-8530, Japan}
\affil{National Astronomical Observatory of Japan, 2-21-1 Osawa, Mitaka, Tokyo 181-8588, Japan}
\author{Hanae Inami}
\affil{Univ Lyon, Univ Lyon1, Ens de Lyon, CNRS, Centre de Recherche Astrophysique de Lyon UMR5574, 69230, Saint-Genis-Laval, France}
\author[0000-0002-8559-6565]{Raffaella A. Marino}
\affil{ETH Z\"urich, Department of Physics, Wolfgang-Pauli-Str. 27, 8093 Z\"urich, Switzerland}
\author[0000-0003-3938-8762]{Sowgat Muzahid}
\affil{Leiden Observatory, Leiden University, P.O. Box 9513, 2300 RA, Leiden, The Netherlands}
\author[0000-0003-2804-0648]{Themiya Nanayakkara}
\affil{Leiden Observatory, Leiden University, P.O. Box 9513, 2300 RA, Leiden, The Netherlands}
\author[0000-0001-5492-1049]{Johan Richard}
\affil{Univ Lyon, Univ Lyon1, Ens de Lyon, CNRS, Centre de Recherche Astrophysique de Lyon UMR5574, 69230, Saint-Genis-Laval, France}
\author[0000-0002-3418-7251]{Kasper B. Schmidt}
\affil{Leibniz-Institut f\"ur Astrophysik Potsdam (AIP), An der Sternwarte 16, 14482 Potsdam, Germany}
\author[0000-0002-2201-1865]{Anne Verhamme}
\affil{Observatoire de Gen\`eve, Universit\'e de Gen\`eve, 51 Ch. des Maillettes, 1290 Versoix, Switzerland}
\author{Lutz Wisotzki}
\affil{Leibniz-Institut f\"ur Astrophysik Potsdam (AIP), An der Sternwarte 16, 14482 Potsdam, Germany}
\begin{abstract}

Using an ultra-deep blind survey with the MUSE integral field spectrograph on the ESO Very Large Telescope, we obtain spectroscopic redshifts to a depth never explored before: galaxies with observed magnitudes $m_{\mathrm{AB}}\gtrsim30$ -- 32.  Specifically, we detect objects via Lyman-$\alpha$ emission at $2.9 < z < 6.7$ without individual continuum counterparts in areas covered by the deepest optical/near-infrared imaging taken by the Hubble Space Telescope, the Hubble Ultra Deep Field.  In total, we find 102 such objects in 9 square arcminutes at these redshifts.  Detailed stacking analyses confirm the Lyman-$\alpha$ emission as well as the 1216 \AA-breaks and faint UV continua (M$_{\mathrm{UV}}\sim -$15).  This makes them the faintest spectroscopically-confirmed objects at these redshifts, similar to the sources believed to reionize the universe.  A simple model for the expected fraction of detected/undetected Lyman-$\alpha$ emitters as a function of luminosity is consistent with these objects being the high-equivalent width tail of the normal Lyman-$\alpha$-emitter population at these redshifts.  
\end{abstract}


\section{INTRODUCTION}

Traditional spectroscopic studies rely on a pre-selection of objects, typically via photometry.  Objects are selected based on a variety of criteria and are then targeted with slits or fibers that feed into a spectrograph.  While this technique is widely used, it is nevertheless unreliable for obtaining complete spectroscopic samples whenever the input photometric catalog is incomplete, e.g. when the objects of interest are near the detection limit of the imaging.  This is particularly true when an emission line is the most significant contribution to the observed broadband magnitude, so objects with spectroscopically-detectable emission lines with high equivalent widths (EWs) might not be present in photometric catalogs \cite[e.g. Figure 7 of][]{Maseda2018}.

Pure blind spectroscopic studies require exquisitely well-understood data to ensure reliable line detections and data from MUSE \cite[the Multi-Unit Spectroscopic Explorer;][]{2010SPIE.7735E..08B}, an Integral Field Spectrograph at the Very Large Telescope, are now in such a state.  The recent survey of the Ultra Deep Field (UDF) with MUSE \cite[][henceforth B17]{Bacon2017} reaches an unprecedented spectroscopic depth ($< 3\times10^{-19}$ \cgs$~$at 7000 \AA; 3-$\sigma$ for a spatially- and spectrally-unresolved line), with quality and uniformity that far exceeds the commissioning data in the Hubble Deep Field South \citep{2015AA...575A..75B}.

One result from B17, which is also hinted at in \citet{2015AA...575A..75B}, is the presence of numerous emission line sources with no counterpart in catalogs based on Hubble Space Telescope (HST) imaging.  While a fraction of these sources are not in photometric catalogs due to close blending issues, the remainder are plausibly extremely faint in the continuum ($m_{\mathrm{AB}} \gtrsim$ 30).  They are believed to be HI \lya emitters (LAEs) at redshifts $2.9 < z < 6.7$ due to an asymmetric line profile and/or the lack of other spectral features which would be indicative of lower-$z$ sources.  The implied ultraviolet (UV) magnitudes (M$_{\mathrm{UV}} > -$16) are intriguing as galaxies this faint are thought to have reionized the universe at $z>6$, but have so far remained elusive spectroscopically \cite[e.g.][]{2012ApJ...752L...5B,2012ApJ...758...93F}.  Detections of the UV continuum or other spectral features would provide further evidence that these ``invisible'' galaxies are indeed high-$z$ LAEs.

In practice, deep non-detections of the UV continuum of an LAE implies that \lya has a large EW.  The interest in these LAEs is due to theoretical expectations where EW$_{\mathrm{Ly\alpha}}$ $\lesssim$ 200 \AA$~$if the photons are produced by normal stellar populations \citep{1993ApJ...415..580C}. This value can be exceeded at extremely low metallicities ($\lesssim 1\%$ Z$_{\odot}$), young ages ($\lesssim$ 10 Myr), or with non-standard stellar initial mass functions, which are potential signatures of the earliest populations of galaxies in the universe \cite[\citeauthor{2003AA...397..527S} \citeyear{2003AA...397..527S}, \citeauthor{2010AA...523A..64R} \citeyear{2010AA...523A..64R}; and also the discussion in][]{2018ApJ...859...53M}.  
While some narrow-band studies have explicitly attempted to constrain the fraction of high-EW LAEs \citep{2002ApJ...565L..71M,2007ApJ...667...79G,2008ApJS..176..301O,2014MNRAS.439.1101Z}, detailed spectroscopic and photometric studies have only confirmed this picture in a few cases \cite[e.g.][]{2012ApJ...761...85K,2017MNRAS.465.1543H}.

Here we present a sample of 102 LAEs detected by MUSE that are not significantly detected in the HST imaging in the UDF, which reaches depths of 29.1 -- 30.3 \citep{XDF}.  We use spectral stacking (Section \ref{sec:undet}) and photometric stacking (Section \ref{sec:beta}) to confirm the MUSE \lya redshifts and estimate the contamination fraction.  Finally, we demonstrate that the observed fraction of HST-undetected LAEs is in line with theoretical expectations (Section \ref{sec:model}).  We adopt a flat $\Lambda$CDM cosmology ($\Omega_m=0.3$, $\Omega_\Lambda=0.7$, and H$_0=70 ~$km s$^{-1}$ Mpc$^{-1}$) and AB magnitudes \citep{1974ApJS...27...21O} throughout.

\section{Data and Sample Selection}

We utilize the MUSE spectroscopic dataset in the UDF, covering 9.92 arcmin$^2$ to 10 hour depth and a single 1.15 arcmin$^2$ subfield to 32 hour depth: further details about the observations and data reduction are presented in B17.  

The positioning of the MUSE data was designed to maximize the overlap with HST imaging in the UDF, the deepest imaging at UV, optical, and near-infrared wavelengths ever taken.  Here, we utilize the reductions from \citet{XDF}, who combine all epochs of imaging from all major surveys in the area (ACS $F435W$, $F606W$, $F775W$, and $F850LP$; WFC3/IR $F105W$, $F125W$, $F140W$, and $F160W$).  We supplement these images with WFC3/UV $F225W$ data \citep{2013AJ....146..159T} and $F275W$ and $F336W$ data \citep{2018arXiv180601853O}.

We use the 160 sources from B17 with flux-weighted emission line centroids that cannot be attributed to photometric objects in the \citeauthor{2015AJ....150...31R} (\citeyear{2015AJ....150...31R}, henceforth R15) catalog within 0\farcs6 \cite[the FWHM of MUSE; see Section 3.1 of][henceforth I17]{Inami2017}.  These sources were identified via a spatially- and spectrally-coherent emission line from the MUSE data using the \texttt{ORIGIN} software (\citeauthor{Bourguignon201232} \citeyear{Bourguignon201232}; \citeauthor{paris:tel-00933827} \citeyear{paris:tel-00933827}; B17; D. Mary et al. in prep.).  As with the full I17 catalog, redshifts are determined via template cross-correlation and human inspection.  Combined with the fact that \lya is spectrally-resolved in MUSE, this means that a majority of the LAEs have fluxes far above the nominal 3-$\sigma$ limit (see Figure \ref{fig:spec} and Table \ref{tab:spec}).

\subsection{Detected versus Undetected}
\label{sec:undet}
With this sample of 160 objects that are not in the R15 catalog, we proceed to measure the magnitudes in order to exclude all HST-detected objects.  We utilize the signal-to-noise (S/N) ratio in HST images within a 0$\farcs$4 aperture centered on the \lya centroid from \texttt{ORIGIN}.  This aperture corresponds to a physical size of 3.1 kpc (2.2 kpc) at $z = 2.9$ (6.6).  However, as shown in Section \ref{sec:beta}, stacked images show that the objects are, on average, more compact than this aperture \cite[and in agreement with measured size-luminosity relations, e.g.][]{2015ApJS..219...15S}.

The local background level is calculated by measuring the standard deviation of the fluxes in 250 identical apertures spread randomly within a 10$''\times$10$''$ cutout centered on the object, with other objects masked according to the R15 segmentation maps.  Because the UDF is not uniform in all photometric bands (specifically the WFC3 coverage), it is crucial to measure the \textit{local} background level instead of relying on the average depth of the field.  If the aperture flux is greater than five times the local background level in an HST band, then we consider the object ``detected'' in that band.  

\begin{figure*}
\begin{center}
\includegraphics[width=.95\textwidth]{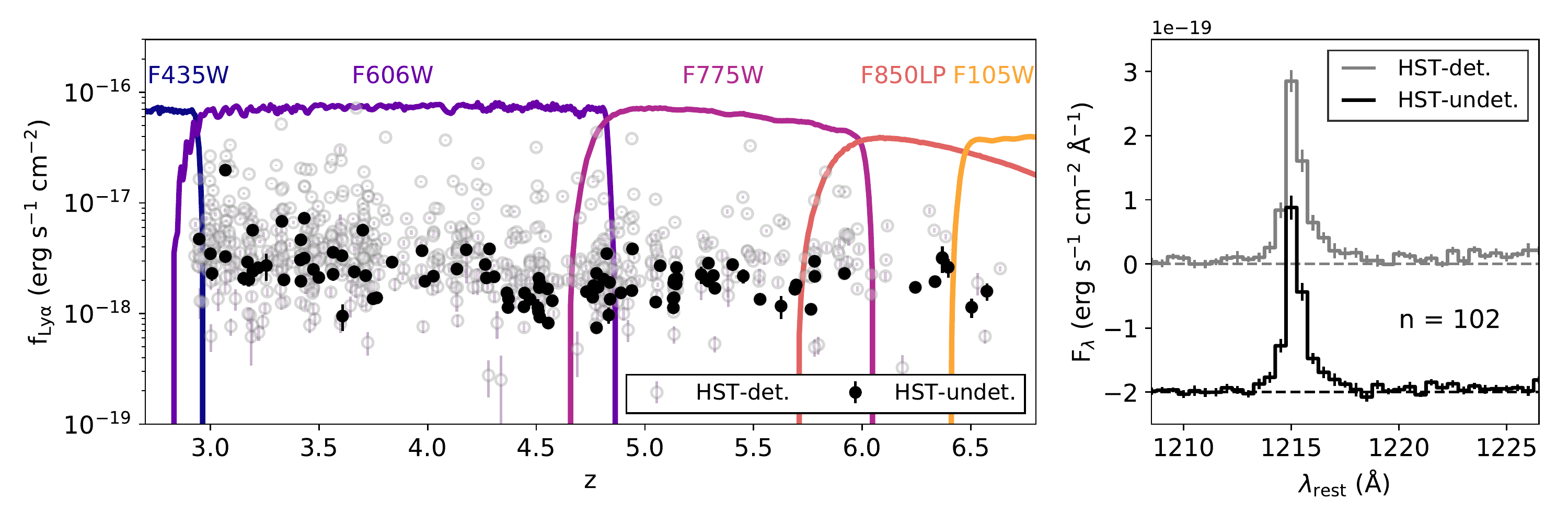}
\end{center}
\vspace{-0.5cm}
\caption{(Left) f$_{\mathrm{Ly\alpha}}$ versus $z$ for the HST-undetected (filled circles) and HST-detected MUSE LAEs (open circles), with HST imaging filter curves denoted by solid lines.  (Right) MUSE spectral stacks (median; bootstrap uncertainties) of LAEs.  The black spectrum shows the HST-undetected LAEs (offset by $-$2) and the gray spectrum is a flux- and redshift-matched sample of 102 HST-detected LAEs.  \lya is clearly detected with good agreement between the HST-detected/undetected LAEs, including the characteristic asymmetry of \lyans.}
\label{fig:spec}
\end{figure*}

In total, we find that 102 of the \texttt{ORIGIN} sources are not detected above 5-$\sigma$ in any of the HST imaging bands, all of which have redshifts classified as \lya (I17).  This represents 12.6\% of the full I17 LAE sample.    At $z > 6$ where an LAE has most of its flux redwards of $F850LP$, we only have a single object that lies outside of the deepest WFC3/IR data in the UDF.

Their \lya fluxes and redshifts compared to the full MUSE sample of HST-detected LAEs are shown in the left panel of Figure \ref{fig:spec}.  Compared to a \lya flux- and redshift-matched sample of HST-detected MUSE LAEs, we see a similar \lya amplitude and spectral profile (right panel of Figure \ref{fig:spec}; see I17 for details on the spectral extractions), confirming the reality of the \texttt{ORIGIN} line detections.  The MUSE ``HST-undetected'' sample is presented in Table \ref{tab:spec}.  The median aperture S/N in $F606W$ of our sample is 1.0, compared with 9.9 for the 663 MUSE-confirmed LAEs that are in the R15  catalog.

\begin{figure}
\begin{center}
\includegraphics[width=.45\textwidth]{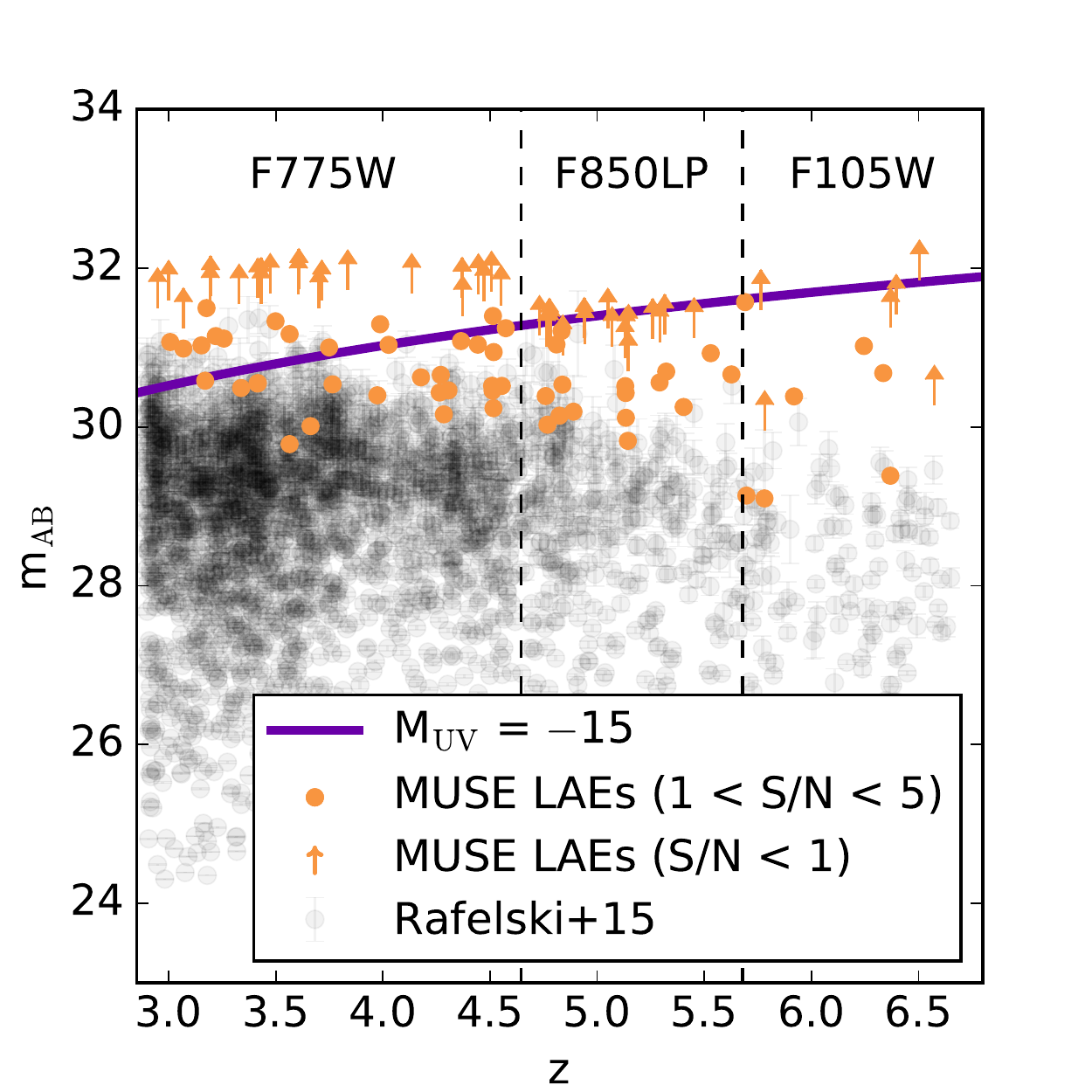}
\end{center}
\caption{Redshift versus observed magnitude for the 102 MUSE HST-undetected LAEs and the photometric sample of R15.  Magnitudes are given in the band immediately redwards of \lyans; for the MUSE objects, the measurement is within a 0\farcs4 aperture, with circles showing measurements (S/N $<$ 5) and arrows showing the 1-$\sigma$ noise level in the aperture when S/N $<$ 1.  The R15 redshifts are photometric whereas the MUSE redshifts are spectroscopic.  A line of constant M$_{\mathrm{UV}}$ = $-$15 is shown in purple, similar to the values from stacks presented in Section \ref{sec:beta}.  The MUSE sample is, by construction, much fainter than the R15 sample.}
\label{fig:zm}
\end{figure}

Of the 58 \texttt{ORIGIN}-only LAEs with an HST detection, 11 are detected only in the photometric band(s) that contains \lyans.  While they are omitted from this sample, they are also plausibly high-EW LAEs since their UV continuum is still undetected (Maseda et al. in prep.).  The remaining sources are not in the R15 catalog primarily because of their projected proximity to brighter galaxies (B17).  Our sample is clearly separated from the HST-detected R15 sources at these redshifts, even though our aperture measurements often represent lower limits to the actual magnitudes (Figure \ref{fig:zm}).

\section{HST Imaging Stacks}

\label{sec:beta}

We create stacks in each imaging band, adopting three redshift bins ($2.974 < z < 4.646$, $4.877 < z < 5.678$, and $6.067 < z < 6.389$) spaced such that the \lya flux lies in a single HST band ($F606W$, $F775W$, and $F850LP$, respectively; see Figure \ref{fig:spec}).  These bins contain 54, 22, and 4 objects.

\begin{figure}
\begin{center}
\includegraphics[width=.45\textwidth]{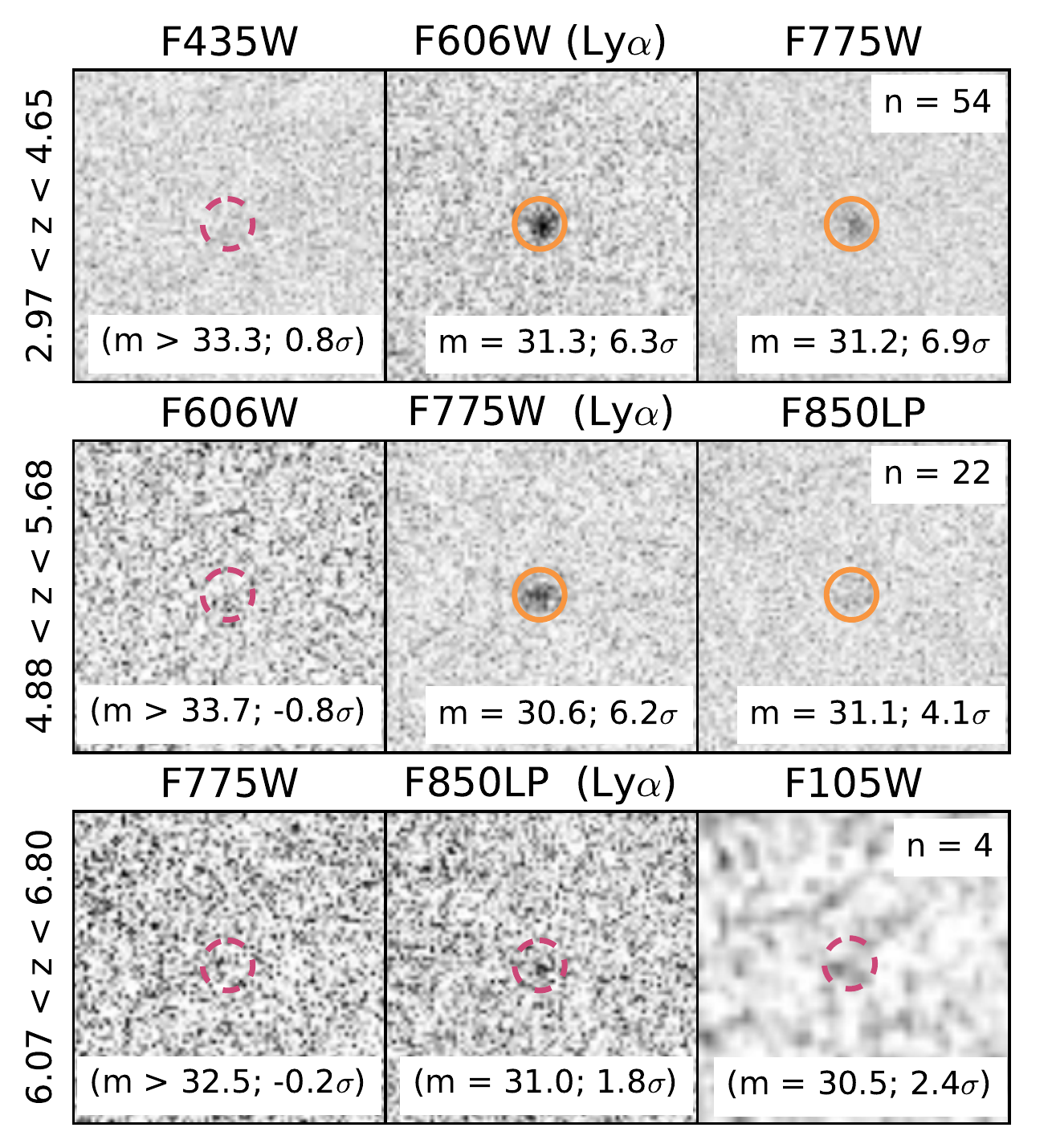}
\end{center}
\caption{HST image stacks for the HST-undetected LAEs, separated into redshift bins.  Each image is 2\farcs5 on a side; the solid (dashed) circle shows a 0\farcs4 aperture with $>$3-$\sigma$ detections (non-detections) in orange (pink). The 1216 \AA-break is demonstrated by non-detections in the left panels as well as \lya emission and the UV continuum in the central/right panels, providing photometric proof that the average object is a high-$z$ LAE.  Numbers of objects per stack, magnitudes (1-$\sigma$ limits when S/N $<$ 1), and detection significances are also shown.}
\label{fig:imstack}
\end{figure}

In each bin we combine the HST imaging for all objects on a filter-by-filter basis.  In each filter stack, we take the mean flux value at each pixel position, using the R15 segmentation maps to mask other sources.  These stacks are shown in Figure \ref{fig:imstack}, restricting the view to the band containing \lya and the bands immediately redwards/bluewards.
The UV continuum appears compact compared to the photometric aperture and therefore we conclude that a majority of our sample have small sizes ($r_e \lesssim$ 1.5 kpc).  In addition, when combining all filters redwards of \lya for \textit{individual} galaxies, 18 have a detection of their UV continuum.  In these cases we do not measure a significant offset between the centroid of \lya and the UV continuum \cite[median 0\farcs09, equal to the size of the HST/ACS point spread function; cf.][]{2011ApJ...735....5F,2017MNRAS.466.1242S}.

We perform aperture photometry on the stacks in the same way as described in Section \ref{sec:undet}.  The UV continuum magnitudes, in the band redwards of \lyans, for the two lower-z bins are $-$14.67 and $-$15.36, and $-$15.37 (1-$\sigma$) for the highest-z bin, at the median redshifts within the bins.  

The effect of \lya on the two lower-$z$ stacks is clear, resulting in strong detections in those bands.  While undetected individually in $F850LP$ and $F105W$, the high-$z$ stack of 4 objects has a significance of 3.0-$\sigma$ when the two filters are combined.  The lack of detections in the blue bands indicates a drop in the spectral energy distribution, likely from the 1216 \AA-break, further demonstrated by the fact that these stacks satisfy the color selections for ``dropout'' galaxies from \citet{2015ApJ...803...34B}.  This all implies that the average galaxy in our sample is indeed at the redshift expected based on the position and identification of \lyans.

If we were to use only the faintest objects in the sample (i.e. S/N $<$ 3 in all HST bands), then we would have stacks of 30, 11, and 4 objects resulting in M$_{\mathrm{UV}}$ values of $-$14.07, $-$15.29, and $-$15.37, also satisfying the ``dropout" color selections.

\subsection{Fraction of Interlopers}

The primary source of contamination in our \texttt{ORIGIN} sample are emission lines being misidentified as \lyans.  \lya is often but not always identified via its characteristic asymmetry, otherwise a single emission feature is identified as \lya when no other emission/absorption features can be detected in the spectrum.  [O II] would be the primary line that is misidentified as \lya considering that other strong optical emission lines (H$\alpha$, [O III], or H$\beta$) are almost never observed alone at MUSE wavelengths and [O II] is the only strong emission line observable at $0.9 < z < 1.5$.  We expect misidentifications predominantly at low S/N, as the spectral resolution of MUSE is high enough to differentiate the two peaks of the [O II] doublet ($\approx$ 200 km s$^{-1}$) from the typical separation of double-peaked $z\approx3$ -- 6 LAEs \cite[$\approx$ 500 km s$^{-1}$:][]{2015ApJ...809...89T,2018MNRAS.tmpL..58V}.

In order to assess the potential contamination fraction we recreate the HST imaging stacks shown in Figure \ref{fig:imstack}, randomly replacing a number of the (plausible) LAEs, $n$, with MUSE-confirmed [O II] emitters with a similar line flux and line position.  For these ``contaminated'' samples with varying $n$ we perform the same stacking procedure, measuring the S/N in the band bluewards of (the misidentified) \lya ($F435W$, $F606W$, and $F775W$, respectively) which is critical in identifying the 1216 \AA-break.  This procedure is repeated 1000 times (replacing a random subset of $n$ objects with [O II] emitters each time) to measure the fraction of cases where the stacks are detected at $>$ 3-$\sigma$.  For these stacks, $n\ge$ 3, 1, and 1 produces detections in the blue band $>$ 99.7\% of the time, implying contamination fractions of $<$ 6\%, $<$ 5\%, and $<$ 25\%, respectively.

\section{Comparison to Continuum-Detected Samples}
\label{sec:model}

In order to assess if the HST-undetected objects are a separate population or an extension of the HST-detected population, we construct an empirical model based on a distribution of rest-frame \lya EWs and a distribution of UV continuum magnitudes.  We use the observed EW distribution for HST-detected MUSE LAEs in the UDF \citep{Hashimoto2017}, assuming no evolution in this distribution with redshift or L$_{\mathrm{Ly\alpha}}$ (see their Section 6.3).  The best-fit lognormal distribution has a mean EW of 119 \AA$~$(15\% of objects have EWs in excess of 200 \AA).  We model the distribution of M$_{\mathrm{UV}}$ values as a power-law with $\alpha$-slopes from \citet{2015ApJ...803...34B} at $z \sim$ 4, 5, and 6.

\begin{figure}
\begin{center}
\includegraphics[width=.4\textwidth]{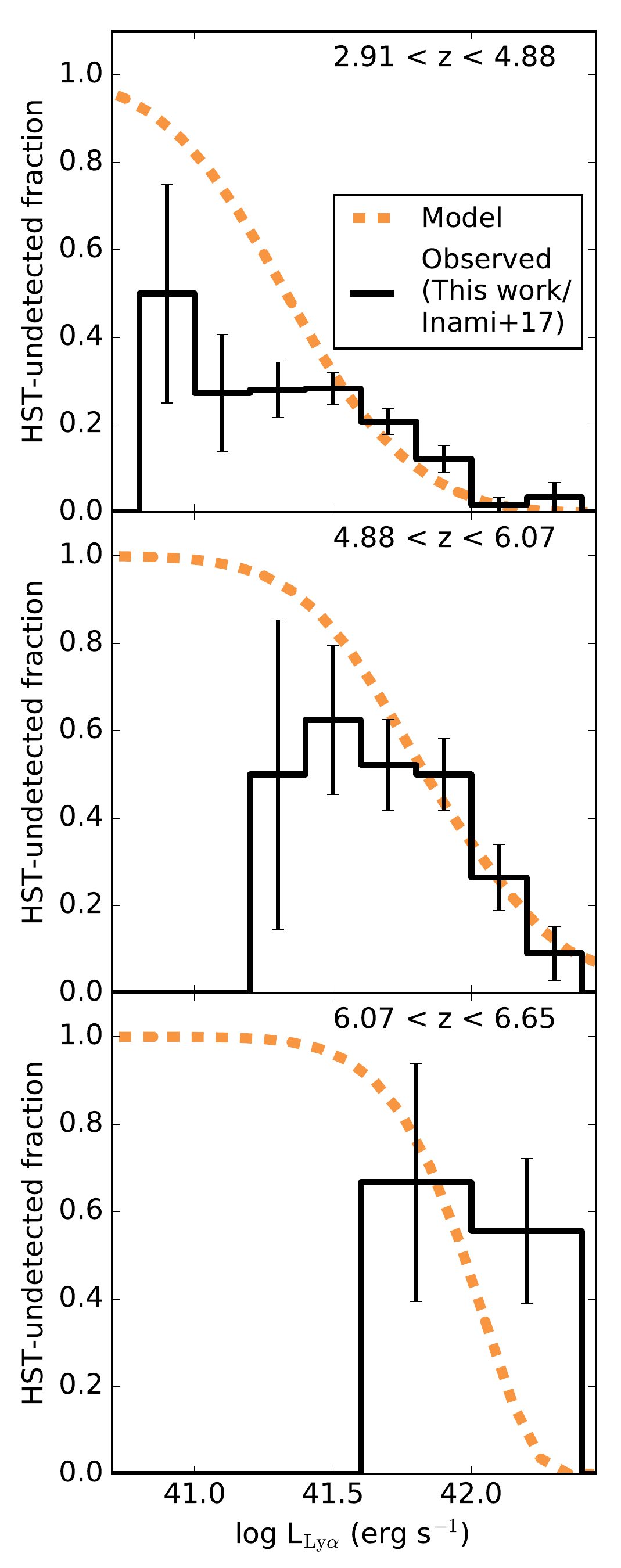}
\end{center}
\caption{Observed (histogram; Poisson errors) and predicted (dashed line) fraction of LAEs in a MUSE-selected sample ($f_{\mathrm{Ly\alpha}} > 3\times10^{-19}$ \cgs) that are undetected in broadband imaging at the depth of the UDF.  The model is based on UV luminosity functions \citep{2015ApJ...803...34B} and a \lya EW distribution \citep{Hashimoto2017}.  Using the observed redshift distribution of undetected sources, the model predicts the dashed curve.  Without any additional tuning we observe good agreement between the prediction and the observed fractions at all redshifts, implying that the HST-detected and undetected objects follow similar trends in UV magnitude and rest-frame \lya EW.}
\label{fig:detfrac}
\end{figure}

By combining these distributions, we can predict the number of objects with a given M$_{\mathrm{UV}}$ and EW$_{\mathrm{Ly\alpha}}$.  These two parameters determine the number of objects that would be spectroscopically-detectable by MUSE with total luminosities (line plus continuum) that would have been observed at the average depth of the HST imaging: 30.1 in $F606W$, 30.1 in $F775W$, and 29.2 in $F850LP$, depending on the redshift \citep{XDF}, and including the mean attenuation of the intergalactic medium \citep{2014MNRAS.442.1805I}.  This model can be inverted for a given L$_{\mathrm{Ly\alpha}}$ to give the number of sources above the HST limits (for the assumed EW distribution).  We use larger redshift ranges ($2.9 < z < 4.88$, $4.88 < z < 6.07$, and $6.07 < z < 6.65$) than in Section \ref{sec:beta} since contribution of \lya to a bluer/redder HST band is not important.  The resulting predictions for the undetected fraction as a function of L$_{\mathrm{Ly\alpha}}$ are shown in Figure \ref{fig:detfrac}, using the observed redshift distribution of the MUSE HST-undetected sources.  Overplotted is the fraction of HST-detected/undetected sources from MUSE.

Overall this model accurately reproduces the observed fractions: comparing the distributions of L$_{\mathrm{Ly\alpha}}$, we obtain $p$-values from a Kolmogorov-Smirnov test of 0.07, 0.68, and 0.97.  While this does not imply with certainty that this is the \textit{best} model to explain the observations, it means that the HST-undetected LAEs are consistent with being the (high-EW) tail  of  the  distribution  of HST-detected  LAEs.

A natural question is whether the phenomenon of a MUSE line detection without an HST counterpart occurs for emission lines other than \lyans.  For example, there are no cases where \texttt{ORIGIN} detects [O II] emission without an HST counterpart.  We can perform a similar analysis to the one above to see if this matches expectations.   We adopt the [O II] luminosity function  from \citet{2013ApJ...772...48P}, measured from HST/ACS slitless grism spectroscopy with coverage of [O II] from $0.9 < z < 1.5$, which is the high-$z$ range probed by MUSE.  At the luminosities probed by MUSE, this function can be approximated by a power law with a slope $\alpha = -1.93$.  This is combined with the EW distribution from \citet{2013ApJ...772...48P} (or I17) to estimate the continuum levels.  We predict that the incidence of HST-undetected [O II] emission is essentially zero for L$_{\mathrm{[O II]}} >$ $10^{39.5}$ erg s$^{-1}$ (6 $\times$ 10$^{-19}$ \cgs$~$at $z=1$).  Only at m$_{\mathrm{F606W}} =$ 29.2 (0.9 magnitudes brighter than our data) would we expect 10\% of $z=1$ [O II] emitters at this luminosity to remain undetected in the continuum.

\section{Discussion and Conclusions}

We have discussed a sample of 102 emission line sources discovered with ultra-deep MUSE spectroscopy in the UDF.  While they are all individually below the detection limits in HST-based imaging, stacks show flux distributions that are expected if these emission lines are predominantly \lyans.  Notably, a strong detection in the HST bands expected to contain \lya and the UV continuum from the two well-populated stacks at $2.974 < z < 4.646$ and $4.877 < z < 5.678$, and in stacked MUSE spectra, implies that the \texttt{ORIGIN} line detections are robust.  A detection in the combined \lya and UV-continuum image in the $6.067 < z < 6.389$ bin also hints at the same conclusion.  We have quantified the amount of contamination from [O II] emission lines that are misidentified as \lya and find that the observed spectral break would disappear for contamination fractions as low as 5\%.  Finally, a simple model utilizing UV luminosity functions and an empirical \lya EW distribution can reproduce the observed fraction of HST-undetected LAEs in our MUSE sample, implying that these sources are consistent with being an extension of the general population of LAEs.

Our stacking experiment reveals M$_{\mathrm{UV}}\sim-$15 for these LAEs, or even $-$14 for the faintest subset.  Compared to spectroscopically-confirmed narrow-band LAEs \cite[e.g.][ M$_{\mathrm{UV}}\lesssim -$17.5]{2008ApJS..176..301O,2011ApJ...734..119K,2014MNRAS.439.1101Z}, Lyman-break galaxies \cite[e.g.][ M$_{\mathrm{UV}}\lesssim -$18]{2010MNRAS.408.1628S}, or MUSE LAEs in the UDF with HST counterparts \cite[][ M$_{\mathrm{UV}}\lesssim -$16]{Hashimoto2017}, this sample is considerably fainter and represents the faintest objects at these redshifts with spectroscopic confirmations.  These magnitudes are comparable to those of local blue compact dwarfs such as I Zw 18 \cite[M$_{\mathrm{UV}}=-14.7$;][]{2007ApJS..173..185G}.  An abundant population of galaxies with such faint magnitudes at $z>6$ are thought to be required in order to reionize the universe \citep{2012ApJ...752L...5B,2012ApJ...758...93F}, yet even our highly sensitive observations confirm $<$ 1\% of the expected numbers at $5 < z < 6.7$ with $-16 < $ M$_{\mathrm{UV}} <  -$14 based on the $z\sim6$ luminosity function \citep{2015ApJ...803...34B}, presumably those with the highest-EWs.

The ability to find emission lines in such faint sources crucially hinges on both the depth of the imaging data (to confirm the faint continua) as well as the depth of the spectroscopic data \cite[these MUSE data probe line fluxes up to 10$\times$ fainter than narrow-band studies at similar redshifts:][]{2017AA...599A..28K, Drake2017, Hashimoto2017}, which is unique to the MUSE UDF Survey.  The next step is to properly characterize the physical properties of this unique population.  By pushing towards LAEs with higher EWs, we can push towards the lowest ages (and hence masses) and metallicities.  While this is challenging with traditional studies \citep{2017MNRAS.465.1543H}, we can perform robust statistical measurements using the MUSE spectroscopic sample due to the stringent constraints provided by the HST imaging.  

\begin{acknowledgements}
We would like to thank the anonymous referee for useful comments that have improved the quality of the manuscript.  RB acknowledges support from ERC grant 339659-MUSICOS.   JB is supported through Investigador FCT contract IF/01654/2014/CP1215/CT0003, national funds (UID/FIS/04434/2013), and by FEDER through COMPETE2020 (POCI-01-0145-FEDER-007672). JS acknowledges support from ERC grant 278594-GasAroundGalaxies. SC acknowledges support from Swiss National Science Foundation grant PP00P2$\_$163824.  TC and NB acknowledge support from ANR FOGHAR (ANR-13-BS05-0010-02), OCEVU Labex (ANR-11-LABX-0060), and the A*MIDEX project (ANR-11-IDEX-0001-02) funded by the ``Investissements d'avenir'' program managed by the ANR.   JR acknowledges support from ERC grant 336736-CALENDS.  AV acknowledges support from ERC grant 757258-TRIPLE.
\end{acknowledgements}

\startlongtable
\begin{deluxetable*}{lllcccccc}
\tablecaption{Properties of HST-undetected MUSE LAEs.\label{tab:spec} (abridged)}
\tablehead{
\colhead{MUSE ID} & \colhead{RA} & \colhead{Dec} & \colhead{z} & \colhead{Ly$\alpha$ Flux} & \colhead{log Ly$\alpha$ Luminosity} & \colhead{S/N $F435W$} & \colhead{S/N $F606W$} & \colhead{S/N $F775W$}  \\
\colhead{(I17)} & \colhead{(deg)} & \colhead{(deg)} & \colhead{} & \colhead{($\times$10$^{-20}$ erg s$^{-1}$ cm$^{-2}$)} & \colhead{(erg s$^{-1}$)} & \colhead{} & \colhead{} & \colhead{} }
\startdata
6316 & 53.17032 & -27.77835 & 4.446 & 153. $\pm$ 7.94 & 41.48 $\pm$ 0.02251 & -1.1 & -0.34 & -1.0\\
6317 & 53.16767 & -27.77743 & 5.404 & 277. $\pm$ 13.6 & 41.94 $\pm$ 0.02140 & 0.23 & 0.58 & 4.2\\
6318 & 53.16665 & -27.77651 & 4.555 & 82.1 $\pm$ 7.49 & 41.23 $\pm$ 0.03963 & -0.29 & 0.022 & 2.9\\
6320 & 53.16465 & -27.78574 & 4.516 & 92.9 $\pm$ 8.20 & 41.28 $\pm$ 0.03831 & -0.0024 & 1.6 & 3.6\\
6321 & 53.16334 & -27.78037 & 3.765 & 139. $\pm$ 13.1 & 41.27 $\pm$ 0.04095 & 0.0049 & 2.7 & 2.7
\enddata
\tablecomments{The unabridged table (102 entries) can be found in the electronic version of the Journal.}
\end{deluxetable*}
\vspace{1.5cm}

\bibliographystyle{aasjournal}

\end{document}